\documentstyle[prl,twocolumn,aps,epsfig]{revtex}
\begin{document}
\draft
\title{Quantum Theory of Atomic Four-Wave Mixing in \\
Bose-Einstein Condensates}
\author{Elena V. Goldstein and Pierre Meystre}
\address{Optical Sciences Center, University of Arizona, Tucson, AZ 85721\\
\medskip}
\author{\small\parbox{14.2cm}{\small\hspace*{3mm}
We present an exact quantum mechanical analysis of collinear four-wave
mixing in a multicomponent Bose-Einstein condensate consisting of sodium
atoms in the $F=1$ ground state. Technically, this is
achieved by taking advantage of the conservation laws of the system to
represent its Hamiltonian in terms of angular momentum operators. We discuss
explicitly the build-up of matter-wave side-modes from noise, as well as
the correlations between these modes. We show the appearance of a strong
quantum entanglement between hyperfine states. We also demonstrate that
for finite atomic numbers, the system exhibits periodic collapses and revivals
in the exchange of atoms between different spin states.
\\[3pt]PACS numbers: 03.75.-b 03.75.Fi 05.30.Jp 42.65.Hw}}
\maketitle
\narrowtext

\section{Introduction}
Low density Bose-Einstein condensates of alkali atoms are described to an
excellent degree of approximation by a Gross-Pitaevskii nonlinear
Schr\"odinger equation. Such equations are ubiquitous in many fields
of physics, including nonlinear optics, hence it is not surprising that
many of the concepts first developed in optics can readily be extended to
condensates. A number of theoretical investigations along these lines
have already been presented, including the study of matter-wave solitons
\cite{Sol},
phase conjugation \cite{GolPlaMey96,GolMey99}, 
four-wave mixing\cite{LawPuBig98}, etc. The experimental verification
of these predictions is also under way, with the first announcement 
of four-wave mixing in a rubidium condensate\cite{Phil98}.

A recent experimental development of considerable importance in this context
is the demonstration of multicomponent condensates, in particular of
$^{23}$Na condensates in far-off-resonance optical dipole traps
\cite{StaAndChi98}. These systems
permit to distinguish condensate modes in ways other than by
their center-of-mass quantum numbers. This opens up new directions of research,
such as the study of the stability and miscibility of quantum fluids, the
analysis of pattern formation, the generation of ferromagnetic and
antiferromagnetic states, etc...
\cite{Ho98,SteInoSta98,LawPuBig98,Phil98}. In the context of
nonlinear atom optics, the coexistence of condensates with
different magnetic quantum numbers is attractive in that it
provides a way to perform four-wave mixing experiments in collinear
geometries, with the considerable advantage of eliminating phase-matching
limitations \cite{GolMey99}.

It is now well understood that the matter-wave analog of  nonlinear optical
interactions is provided by collisions. In particular, 
two-body collisions in the
shapeless approximation are mathematically equivalent to a local Kerr medium
with instantaneous response. Hence, the kinds of wave-mixing phenomena that
can take place in a condensate are largely dictated by the properties of
these collisions. For example, recent experimental studies of spin
population dynamics in a $^{23}$Na condensate have shown that as a
result of spin changing collisions, a sample initially in the $m_F = 0$ state
eventually winds up with almost equal populations of all hyperfine ground
states \cite{SteInoSta98}. 
The present paper exploits these collisions to generate
four-wave mixing in $^{23}$Na. It is organized as follows: Section II
describes our physical model, which considers degenerate backward phase
conjugation in (small) condensates. Section III derives a representation of the
problem in terms of angular momentum algebra. This leads to an exact
diagonalization of the four-wave mixing Hamiltonian and the determination of
all eigenstates and eigenenergies. These results are applied in Section IV
to the analysis of the dynamics of exchange of population between different
spin states. The quantum correlations between these states are also
discussed. Finally, Section V is a summary and outlook.

The phase conjugation of atomic waves has previously been discussed in
a situation where the matter-wave modes involve the same electronic
state, but different center-of-mass components, and using the undepleted
pump approximation \cite{GolPlaMey96}. 
Mode separation was achieved via Bragg scattering
off an optical field. A full three-dimensional numerical solution of a
closely related situation was recently given \cite{TriBanJul98}
in connection with experimental work at NIST \cite{Phil98}. 
Both of these analyses rely on the
Gross-Pitaevskii equation, which effectively assumes a Hartree ansatz
with a fixed number of particles, or spontaneous symmetry breaking.
In addition, a recent analysis of matter-wave phase conjugation treats the
central mode to all orders, but describes the side-modes in a linearized
fashion \cite{GolMey99}.
In contrast, the present theory, which also holds for small condensates,
does not make any of these assumptions. It handles all modes on equal footing,
to all orders, and makes no assumption about their statistical properties.
We note that our approach is closely related in spirit to a recent analysis
of matter-wave four-wave mixing \cite{LawPuBig98}, which
is also valid to all orders and uses angular momentum algebra to
diagonalize the problem.

\section{Physical model}

We consider a Bose-Einstein condensate of $^{23}$Na atoms in the $F=1$
hyperfine ground state, with the three internal atomic states 
$|F=1,m_F=-1\rangle$,
$|F=1,m_F=0\rangle$ and $|F=1,m_F=1\rangle$ of degenerate energies in the
absence of external magnetic fields. The condensate is confined 
by a far-off-resonance optical dipole trap. It is described by the 
three-component vector Schr\"odinger field
\begin{equation}
\bbox{\Psi}({\bf r},t)
=\{\Psi_{-1}({\bf r},t),\Psi_{0}({\bf r},t),\Psi_{1}({\bf r},t)\}
\end{equation}
which satisfies the bosonic commutation relations
\begin{equation}
[\Psi_i({\bf r},t), \Psi_j^\dagger({\bf r}',t)]
=\delta_{ij}\delta({\bf r}-{\bf r}').
\end{equation}
Accounting for the possibility of two-body collisions,
its dynamics is described by the second-quantized Hamiltonian
\begin{eqnarray}
& &{\cal H}=\int d {\bf r} \bbox{\Psi}^\dagger({\bf r},t)H_0
\bbox{\Psi}({\bf r},t)
\nonumber\\
& &+\frac{1}{2}\int d {\bf r}_1d{\bf r}_2\bbox{\Psi}^\dagger({\bf r}_1,t)
\bbox{\Psi}^\dagger({\bf r_2},t)V({\bf r}_1-{\bf r}_2)
\bbox{\Psi}({\bf r}_2,t)\bbox{\Psi}({\bf r}_1,t),
\label{ham2}
\end{eqnarray}
where the single-particle Hamiltonian is given by
\begin{equation}
H_0={\bf p}^2/2M + V_{trap}
\end{equation}
and the trap potential is of the general form
\begin{equation}
V_{trap} = \sum_{m=-1}^{+1} U({\bf r})|F=1,m\rangle\langle F=1,m|.
\end{equation}
Here ${\bf p}$ is the center-of-mass momentum of the atoms of mass $M$ and
$U({\bf r})$, the  effective dipole trap potential for
atoms in the $|1,m \rangle$ hyperfine state, is independent of
$m$ for a non-magnetic trap.

The general form of the two-body interaction $V({\bf r}_1-{\bf r}_2)$
has been discussed in detail in Refs. \cite{Ho98,ZhaWal98}. We reproduce
its main features for the sake of clarity. Considering exclusively 
situations where the hyperfine spin $F_i = 1$ of the individual atoms is 
preserved, we label the hyperfine states of the combined system of
two collision partners with total hyperfine spin ${\bf F} = {\bf F}_1 +
{\bf F}_2$ by $|f,m \rangle$, where $f = 0, 1,2$ and $m=-f,\ldots,f$.
In the shapeless approximation, it can be shown that the two-body
interaction is of the general form \cite{Ho98}
\begin{equation}
 V({\bf r}_1-{\bf r}_2)=\delta({\bf r}_1-{\bf r}_2)\sum_{f=0}^{2}
\hbar g_f {\cal P}_f,
\label{pot}
\end{equation}
where
\begin{equation}
g_f=4\pi\hbar a_f/M,
\end{equation}
and
\begin{equation}
{\cal P}_f\equiv \sum_m|f,m\rangle\langle f,m|
\end{equation}
is the operator which projects the state of the atomic pair onto
a state of total hyperfine quantum number $f$. Here $a_f$ is the $s$-wave
scattering length for the channel of total hyperfine spin $f$.
As a result of the symmetry requirement 
for bosonic atoms, it can be shown that
only states with even $f$ contribute to $ V({\bf r}_1-{\bf r}_2)$, so 
that
\begin{eqnarray}
V({\bf r}_1-{\bf r}_2)&=&\hbar \delta({\bf r}_1-{\bf r}_2)(g_2{\cal P}_2+
g_0{\cal P}_0) \nonumber\\
&=&\hbar\delta({\bf r}_1-{\bf r}_2)
\left(c_0+c_2{\bf F}_1\cdot{\bf F}_2 \right).
\end{eqnarray}
In this expression,
\begin{eqnarray}
c_0 &=& (g_0+2g_2)/3 \nonumber \\
c_2 &=& (g_2-g_0)/3,
\end{eqnarray}
which follows from the identities ${\cal P}_1+{\cal P}_2={\hat I}$ and
${\bf F}_1\cdot{\bf F}_2={\cal P}_2-2{\cal P}_0$. Substituting this form of 
the two-body potential $V({\bf r}_1-{\bf r}_2)$ into the second-quantized
Hamiltonian (\ref{ham2}) leads to
\begin{eqnarray}
& &{\cal H}=\sum_m\int d{\bf r} \Psi_m^\dagger({\bf r},t)
\left[\frac {{\bf p}^2}{2M}+U({\bf r})\right]\Psi_m({\bf r},t)\nonumber\\
& &
+\frac{\hbar}{2}\int d{\bf r}\{(c_0+c_2)
[\Psi_1^\dagger\Psi_1^\dagger\Psi_1\Psi_1
+\Psi_{-1}^\dagger\Psi_{-1}^\dagger\Psi_{-1}\Psi_{-1}
\nonumber\\
& &
+2\Psi_0^\dagger\Psi_0
(\Psi_1^\dagger\Psi_1+\Psi_{-1}^\dagger\Psi_{-1})]
+c_0\Psi_0^\dagger\Psi_0^\dagger\Psi_0\Psi_0
\nonumber\\
& &
+2(c_0-c_2)\Psi_1^\dagger\Psi_1
\Psi_{-1}^\dagger\Psi_{-1}
\nonumber\\
& &
+2c_2(\Psi_1^\dagger\Psi_{-1}^\dagger\Psi_0\Psi_0+H.c. )\}.
\label{ham22}
\end{eqnarray}

The physical interpretation of the various terms of this Hamiltonian
has been discussed previously: The three terms quartic in one of the 
field operators, i.e. those of the form 
$\Psi_i{^\dagger} \Psi_i^{\dagger} \Psi_i \Psi_i$
are self-(de)focussing terms, the terms involving two hyperfine states
conserve the populations of the individual spin states and merely lead to
phase shifts, and the terms involving the central mode $\Psi_0$ and
{\em both} side-modes correspond to spin-exchange collisions. This
``four-wave mixing'' interaction, involving e.g. the annihilation of a pair
of atoms with $m_F = 0$ and the creation of two atoms in the states
$m_F = \pm 1$,  leads to phase conjugation in quantum optics and to
matter-wave phase conjugation in the present case \cite{GolMey99}.

The analogy with optical four-wave mixing becomes even more apparent when
we consider a situation where atoms in the $m_F=0$ state are placed in
a linear superposition of two counterpropagating center-of-mass
modes of momenta $\pm \hbar k_0$, that is
\begin{equation}
\Psi_0(x)=\frac{1}{\sqrt V}\left(e^{ik_0 x} a_{01}+e^{-ik_0 x}a_{02}\right),
\label{cond}
\end{equation}
while the atoms of spin $m_F = \pm 1$ are taken to be at rest
\begin{equation}
\Psi_{\pm1}(x) = \frac{1}{\sqrt V}a_{\pm 1}.
\label{side}
\end{equation}
Here $a_{01}$  and $a_{02}$ are the annihilation operators of the two
counterpropagating $m_F = 0$ modes, with $[a_{0i}, a^\dag_{0j}] =
\delta_{ij}$, $i, j = 1$ or 2, and
$a_{1},  a_{-1}$ are the corresponding operators for the modes
associated with $m_F = \pm 1$. Finally, $V$ is the is the confinement
volume of the condensate. Inserting this mode expansion into the
Hamiltonian (\ref{ham22}) and ignoring all non-phase matched contributions
finally yields the four-wave mixing Hamiltonian
\begin{eqnarray}
{\cal H}&=&\frac{\hbar^2k_0^2}{2M}\hat N+ \frac{\hbar c_0}{2} 
{\hat N}({\hat N}-1)
\nonumber\\
&+&\frac{\hbar c_2}{2}
(a_1^\dagger a_1^\dagger a_1 a_1+a_{-1}^\dagger a_{-1}^\dagger
a_{-1} a_{-1}-2 a_1^\dagger a_1 a_{-1}^\dagger a_{-1}
\nonumber\\
&+&2 (a_1^\dagger a_1+ a_{-1}^\dagger a_{-1})
(a_{01}^\dagger a_{01}+a_{02}^\dagger a_{02})
\nonumber\\
&+&4a_1^\dagger a_{-1}^\dagger a_{01}a_{02}+4 a_1a_{-1}a_{01}^\dagger
a_{02}^\dagger),
\label{modham}
\end{eqnarray}
where we have introduced the total number of atoms
\begin{equation}
{\hat N}=a_1^\dagger a_1+a_{-1}^\dagger a_{-1}+a_{01}^\dagger a_{01}
+a_{02}^\dagger a_{02}.
\label{total}
\end{equation}
Note that the coupling between the
central and side-modes of the condensate
requires both energy and momentum conservation. 
Energy conservation can be achieved e.g. by introducing an external dc 
magnetic field. Then the quadratic Zeeman effect compensates the mismatch 
in kinetic energies and provides the necessary 
condition $E_1+E_{-1}=\hbar^2 k_0^2/m$ ($E_\pm 1$ being the energies
of $m_F=\pm 1$ spin states ). Indeed, for the magnetic
fields used in Ref. \cite{SteInoSta98} $E_1+E_{-1}\sim 10^{-28}J$ which is of
the same order as the central mode kinetic energy $\hbar^2 k_0^2/2m$
for $k_0\sim 10^7$m.

\section{Angular momentum representation}

Despite the formal analogy between the Hamiltonians describing optical
and atomic four-wave mixing, there are important details in which these
processes differ. One of them is phase matching, which is normally more
difficult to achieve with de Broglie waves due to the quadratic dispersion
relation of atoms. However, the use of multicomponent condensates permits
to avoid this difficulty, as we have just seen. More important perhaps is
the fact that nonlinear optics experiments usually use coherent laser light
as pumps. It is well known that the number of photons in these fields is
not well determined. In contrast, the number of atoms in a condensate is
a fixed (and integer) quantity in the absence of losses. This leads 
to important
conservation laws, that we exploit in the following to reexpress the
Hamiltonian (\ref{modham}) in terms of angular momentum operators. This
leads in turn to an exact solution of the problem.

The most obvious conserved quantity of the Hamiltonian (\ref{total}) is
the total number of atoms, as follows from the commutator
\begin{equation}
[{\hat N},{\cal H}]=0
\end{equation}
In addition, the population {\em differences}
\begin{eqnarray}
{\hat D}_0&=&a_{01}^\dagger a_{01}-a_{02}^\dagger a_{02}\\
{\hat D}&=&a_{1}^\dagger a_{1}-a_{2}^\dagger a_{2}
\label{diff}
\end{eqnarray}
are also conserved. These two conservation laws follow directly from the
fact that the Hamiltonian (\ref{modham}) describes the creation and
annihilation of bosonic atoms in pairs. For example, the annihilation
of two atoms in the center-of-mass modes 1 and 2 of the $m_F = 0$ hyperfine
state, described by the operator pair $a_{01}a_{02}$,
results in the creation of a pair of atoms in the $m_F = \pm 1$ spin states
via $a_1^\dagger a_{-1}^\dagger$.

Together with the conservation of the total number of atoms, 
these conservation laws yield the two additional conserved quantities
\begin{eqnarray}
{\hat N}_1&\equiv& a_1^\dagger a_1+a_{01}^\dagger a_{01}\nonumber\\
\mbox{and }\nonumber\\
{\hat N}_2&\equiv& a_{-1}^\dagger a_{-1}+a_{02}^\dagger a_{02}.
\label{total2}
\end{eqnarray}

These conservation laws make it possible to define an angular momentum
algebra for this four-wave mixing system analogous to the
Schwinger coupled boson representation used in Ref. \cite{MilCorWri97}
for the description of two-mode condensates and in Ref. \cite{LawPuBig98}
for the three-mode coupling problem. However the present situation
requires that these considerations be extended to  
a compound angular momentum representation \cite{Mat81}.

We proceed by introducing spinor operators ${\bf a}_1$ and ${\bf a}_2$,
\begin{equation}
{\bf a}_1\equiv\pmatrix{
 a_{01}\cr
a_1} \mbox{\hspace{2.mm} and \hspace{2.mm}}
{\bf a}_2\equiv\pmatrix{
 a_{2}\cr
a_{02}},
\label{spinor}
\end{equation}
as well as the two angular momentum operators ${\bf S}_1$ and 
${\bf S}_2$,
\begin{equation}
{\bf S}_1\equiv{\bf a}_1^\dagger {\bbox \sigma}{\bf a}_1
=\frac{1}{2}\pmatrix{a_{01}^\dagger a_1+a_{01}a_1^\dagger\cr
-{i}(a_{01}^\dagger a_1-a_{01}a_1^\dagger) \cr
a_{01}^\dagger a_{01}-a_1^\dagger a_1}
\label{angmom1}
\end{equation}
and
\begin{equation}
{\bf S}_2\equiv{\bf a}_2^\dagger {\bbox \sigma}{\bf a}_2
=\frac{1}{2}\pmatrix{a_{2}^\dagger a_{02}+a_{02}^\dagger a_2 \cr
-{i}(a_{02} a_2^\dagger -a_{02}^\dagger a_2)\cr
a_{2}^\dagger a_{2}-a_{02}^\dagger a_{02}},
\label{angmom2}
\end{equation}
where $\bbox {\sigma}$ is the Pauli spin operator.
The Casimir operators $K_j$ associated with the angular momenta 
${\bf S}_j$ are also constants of motion since
\begin{equation}
K_j\equiv  {\bf S}_j^2=
\frac{{\hat N}_j}{2}\left(\frac{{\hat N}_j}{2}+1\right).
\end{equation}

Expressed in terms of the total spin operator
\begin{equation}
{\bf S}\equiv {\bf S}_1+{\bf S}_2
\end{equation}
with components ${ S}_j={ S}_{1j}+{ S}_{2j}$, 
the Hamiltonian (\ref{modham}) becomes
\begin{eqnarray}
{\cal H}&=&  \frac{\hbar^2 k_0^2}{2M}{\hat N} + 
\frac{\hbar c_0}{2}{\hat N}({\hat N}-1)\nonumber\\
&+& {2\hbar c_2}\left[{\bf S}^2-\frac{\hat N}{2}
-\left(\frac{\hat{D}_0}{2}\right)^2\right].
\label{spinham}
\end{eqnarray}
Together with the conserved Casimir operators, this form of the four-wave
mixing Hamiltonian suggests expressing the state of the system in terms of
the complete set of quantum numbers associated with the eigenstates of the
operators ${\bf S}_1^2$, ${\bf S}_2^2$, ${\bf S}^2$
and $S_z$. The Hamiltonian (\ref{spinham}) can clearly be diagonalized
in terms of these states, yielding a complete set of eigenstates and
eigenenergies.

To illustrate how this works, we consider for concreteness a condensate
consisting of $N$ atoms such that ititially
$N_1 = N_2 = N/2$, $m \ll N/2$ atoms in the
hyperfine state $m_F = 1$ and none in the state $m_F = 0$. In the
``natural'' eigenbasis of the operators $\{K_1, S_{1z}, K_2,
S_{2z}\}$
the initial state of this system is described by the state vector
\begin{equation}
|\phi(0)\rangle =|\frac{N}{4} ,\frac{N}{4}-m;
\frac{N}{4},-\frac{N}{4}\rangle.
\end{equation}
Expanding it in terms of a complete set of eigenvectors
$\{|\frac{N}{4}, \frac{N}{4}, S, S_z \rangle \}$ of the Hamiltonian
(\ref{spinham}), we find
\begin{equation}
|\phi(0)\rangle =
\sum_{S} {\cal C}(\frac{N}{4}-m,-\frac{N}{4};S,-m)
|\frac{N}{4},\frac{N}{4};S,-m\rangle,
\end{equation}
where
\begin{equation}
{\cal C}(S_{1z},S_{2z};S,-m)\equiv
\langle \frac{N}{4},\frac{N}{4};S,-m|\frac{N}{4},S_{1z};\frac{N}{4},S_{2z}
\rangle
\end{equation}
are Clebsch-Gordan coefficients \cite{AbrSte72}, which are nonzero only
for $S_z=S_{1z}+S_{2z}=-m$.

This conservation of $S_z$ under the Hamiltonian (\ref{spinham})
further implies that the state vector of the system is given at time $t$ by
\begin{eqnarray}
|\phi(t)\rangle&=& \sum_{S}\alpha_S(t)|\frac{N}{4},\frac{N}{4};S,-m\rangle
\nonumber\\
&\equiv&\sum_{S}\alpha_S(t)|S,-m\rangle,
\label{wf}
\end{eqnarray}
where  in the second equality we have made
explicit use of the value of the conserved quantities $K_1$, $K_2$
and $S_z$ to simplify the notation via
\begin{equation}
|\frac{N}{4},\frac{N}{4};S,-m\rangle \rightarrow |S,-m \rangle .
\label{simple}
\end{equation}
Note that this simplification is not general, but is appropriate for the
initial condition at hand.

The equations of motion for the probability amplitudes
$\alpha_S(t)$
follow from the Schr\"odinger equation. For the condensate containing $N$
atoms they read, in a frame rotating at the frequency 
$\hbar k_0^2N/2M + c_0N(N-1)/2-2 c_2(N/2+(D_0/2)^2)$,
\begin{equation}
i\dot \alpha_S(t)=2c_2S(S+1) \alpha_S(t)
\end{equation}
and can be integrated trivially with the initial condition
$$\alpha_S(t=0)={\cal C}(\frac{N}{4}-m,-\frac{N}{4};S,-m).$$

\section{Dynamics}

First we apply the results of the preceding section to the study of the
dynamics of population exchange between the different modes of the
condensate. For instance, the population of the $m_F = 1$ hyperfine spin
state can readily be determined from the expectation
value of $S_{1z}$. With Eqs. (\ref{total2})
and the definition of $S_{1z}$ we find
\begin{equation}
\langle a_1^\dagger a_1\rangle=\frac{N}{4}-\langle S_{1z}\rangle.
\end{equation}
From Eq. (\ref{wf}) we have
\begin{equation}
\langle S_{1z}\rangle
=\sum_p p \left |\sum_S \alpha_S(t)
{\cal C}(p,-(p+m);S,-m) \right|^2,
\label{pop1}
\end{equation}
where we inserted the identity operator
\begin{equation}
{\hat I} = \sum_{p_1,p_2}|\frac{N}{4},p_1;\frac{N}{4},p_2\rangle
\langle \frac{N}{4},p_1;\frac{N}{4},p_2|
\end{equation}
and used the simplified notation (\ref{simple}) as well as the property that
the Clebsch-Gordan coefficients' are nonzero only for
$S_z=S_{1z}+S_{2z}$. Eq. (\ref{pop1}) can be evaluated numerically.

The evolution of the population of the $m_F = 1$ sidemode is shown in Fig. 1
for $N=100$ atoms in the system. In case (b) the initial mode population
is $\langle a_1^\dagger a_1 \rangle = m = 5$, while case (a) illustrates
the build-up from noise, $m = 0$. In both cases, the sidemode population
exhibits an initial growth to the point where it contains about $1/3$ of the
atoms in the first case and about half of the atoms in the second. This
is followed by a collapse to a quasi-steady state population, as well as
a subsequent revival at $2c_2t_1=\pi$. This dynamics then repeats
itself periodically, with revivals at $2c_2t_n=\pi n$, independently of
$N$. This is similar to the periodical revivals which occur in two-photon
Janes-Cummings model discussed in \cite{BucSuk81}.

During the periods of collapse, one has $S_{jz}=-m/2$, so that all modes
are almost equally macroscopically populated with
$\langle a_{01}^\dagger  a_{01}\rangle=N/4-m/2$,
$\langle a_{1}^\dagger  a_{1}\rangle=N/4+m/2$,
$\langle a_{02}^\dagger  a_{02}\rangle=N/4+m/2$ and
$\langle a_{2}^\dagger  a_{2}\rangle=N/4-m/2$.

A particularly interesting aspect of the present study is that it allows one
to obtain the quantum correlations between sidemodes. In optics, for example,
four-wave mixing provides a method to study purely quantum mechanical effects 
such as squeezing and nonclassical states of the radiation field, and also 
to prepare states of composite systems exhibiting strong quantum mechanical
entanglement \cite{WalMil94}. 
These states are of considerable interest in tests of the 
foundations of physics as well as quantum information processing such as 
quantum cryptography \cite{Ben92,TitRibGis98}
and quantum computing \cite{EkeJoz96}. Macroscopic quantum states
of massive particles present an interesting alternative to all-optical
systems, hence it is of considerable interest to determine to which extent
quantum entanglement between sidemodes can be achieved in Bose-Einstein
condensation.

In analogy with the optical case, one can quantify the amount of quantum
entanglement between condensate modes by determining the extent to which
the Cauchy-Schwartz inequality is violated by the second-order
cross-correlation functions between modes \cite{WalMil94}. In particular, 
for a 'classically looking ' optical system with
positive Glauber ${\cal P}$-representation,  the single-time 
second-order cross-correlation function is bound by 
\begin{equation}
G^{(2)}_{i,j}(t) \leq
[G^{(2)}_{i}(t)G^{(2)}_{j}(t)]^{1/2} .
\end{equation}
In case the ${\cal P}$-representation is not positive or
does not exist, in contrast, the upper bound is higher, namely
\begin{equation}
G^{(2)}_{i,j}(t)\leq\left\{\left[G^{(2)}_{i}(t)
+G^{(1)}_{i}(t) \right)
\left(G^{(2)}_{j}(t)+G^{(1)}_{j}(t)\right]\right\}^{1/2}.
\label{CSQ}
\end{equation}
In these inequalities, we have introduced the single-time and single-mode 
first-order correlation functions
\begin{equation}
G^{(1)}_{j}(t)\equiv \langle \phi(t)|a_j^\dagger a_j|\phi(t)\rangle
\end{equation}
as well as the single-time two-mode second-order correlation functions
\begin{equation}
G^{(2)}_{ij}(t)\equiv{\langle \phi(t)|a_i^\dagger a_ia_{j}^\dagger
a_{j}|\phi(t)\rangle}
\end{equation}
and the single-time, second-order correlation functions
\begin{equation}
G^{(2)}_{j}(t)\equiv{\langle \phi(t)|a_j^\dagger a_j^\dagger
a_{j} a_{j}|\phi(t)\rangle} .
\end{equation}

As was the case for the sidemode populations, the single-time single-mode
second-order cross-correlation between the sidemodes $m_F =
\pm 1$ can be expressed in terms of the $z$-component of the individual 
pseudospins as
\begin{eqnarray}
& &G^{(2)}_{-1,1}(t)=
\langle (\frac{N}{4}-S_{1z})(\frac{N}{4}-S_{2z})\rangle
\nonumber\\
& &=\left (\frac{N}{4}\right)^2-\frac{N}{4}\left(\langle S_{1z}\rangle
+\langle S_{2z}\rangle\right)+\langle S_{1z}S_{2z}\rangle,
\label{spincor}
\end{eqnarray}
where
\begin{eqnarray}
&&\langle S_{1z}S_{2z}\rangle \nonumber \\
&&=\sum_p p(-p-m) \left |\sum_{s=m}^{[N/2]}
\alpha_S(t){\cal C}(p,-(p+m);S,m)\right|^2
\end{eqnarray}
and the sum can easily be evaluated numerically.

Figure 2 compares the time dependence of the normalized 
central-mode---side-mode
correlation function (lower curve)
\begin{equation}
{\cal R}^{(2)}_{01,1}(t)\equiv \frac {G^{(2)}_{01,1}(t)}
{ \sqrt{ G^{(2)}_{01}(t)G^{(2)}_{1}(t)}}
\end{equation}
and the side-mode---side-mode correlation (upper curve)
\begin{equation}
{\cal R}^{(2)}_{-1,1}(t)\equiv
\frac {G^{(2)}_{-1,1}(t)}{\sqrt{G^{(2)}_{1}(t)G^{(2)}_{-1}(t)}}
\end{equation}
for (a) the case $m=0$ where the sidemode builds up from quantum fluctuations
and (b) the case $m=5$ of an injected signal. These results illustrate how
the correlations between central-mode  and side-modes do satisfy the
classical Cauchy-Schwartz inequality while the side-mode---side-mode
cross correlations violate them. The violation is particularly strong in the
case of build-up from noise, as should be intuitively expected. In that case
the hyperfine sidemodes $m_F = \pm1$ play symmetric roles, thus
\begin{equation}
G^{(2)}_{-1,1}(t)=G^{(2)}_{1}(t)+G^{(1)}_{1}(t),
\end{equation}
i.e. Eq. (\ref{CSQ}) thus becomes an
equality, corresponding to the maximum violation of the 
classical Cauchy-Schwartz inequality allowed by quantum mechanics.

The difference in the behavior of the two-mode correlation functions between
side-modes and those involving one central and one side-mode can be 
intuitively understood from the form of the
wave-mixing term $a_1^\dagger a_{-1}^\dagger a_{01}a_{02}$ appearing in the
Hamiltonian (\ref{modham}). Indeed, the coupling between side-modes, involving
two annihilation operators, is reminiscent of the interaction 
$a_1^\dagger a_2^\dagger$ in the Hamiltonian of parametric amplification 
leading to squeezing and quantum entanglement between two sidemodes. 
In contrast, the coupling between central and sidemodes involves both an
annihilation and a creation operator.

\section{Summary and conclusions}

In contrast to the optical case, where it is usually difficult to
study quantum mechanically four-wave mixing in the strong field limit, this
task can be achieved relatively easily for matter waves, a direct consequence
of the conservation of the total number of atoms in a condensate at $T=0$ and
in a lossless trap. The resulting conservation laws permit to develop an
angular momentum algebra analysis that leads to an exact solution of the
problem away from the linear regime where the sidemode populations 
remain small.

In this paper we have applied this technique to the study of the
dynamics of the population exchange between hyperfine levels of a condensate.
We found that this exchange is periodic and it is characterized by a sequence
of ``collapses'' and ``revivals'' reminiscent of those appearing in the
two-photon Jaynes-Cummings model of optical physics. In addition, we found
that strong quantum correlations can develop between the central modes and
the side modes, the general state of the system exhibiting a strong quantum
entanglement between the modes $m_F = \pm 1$. Thus, it appears that
multicomponent condensates offer a fascinating method to create
quantum entanglement at a truly macroscopic level, a possibility made
even more attractive by the fact that these systems suffer very little
from dissipation, since they consist of ground-state atoms. The finite
lifetime of condensates is usually attributed to three-body collisions, which
result in losses on a timescale of seconds.

We conclude by noting that the revivals in the population exchange occur at
a time independent of the number of atoms in the condensate. Hence, they
allow for a direct and absolute determination of the coefficient $c_2$.
In practice, however, $c_2$ seems to be too small
to observe revivals on the timescales of the condensate lifetime.

\acknowledgements
This work is supported in part by the U.S. Office of Naval Research 
Contract No. 14-91-J1205, by the National Science Foundation Grant No.
PHY95-07639, by the U.S. Army Research Office and by the
Joint Services Optics Program. The valuable comments and suggestions 
by O. Zobay are greatfully acknowledged. 
E. V. G. also acknowledges stimulating discussions with J. H. Eberly.


\newpage
\begin{figure}
\epsfig{figure=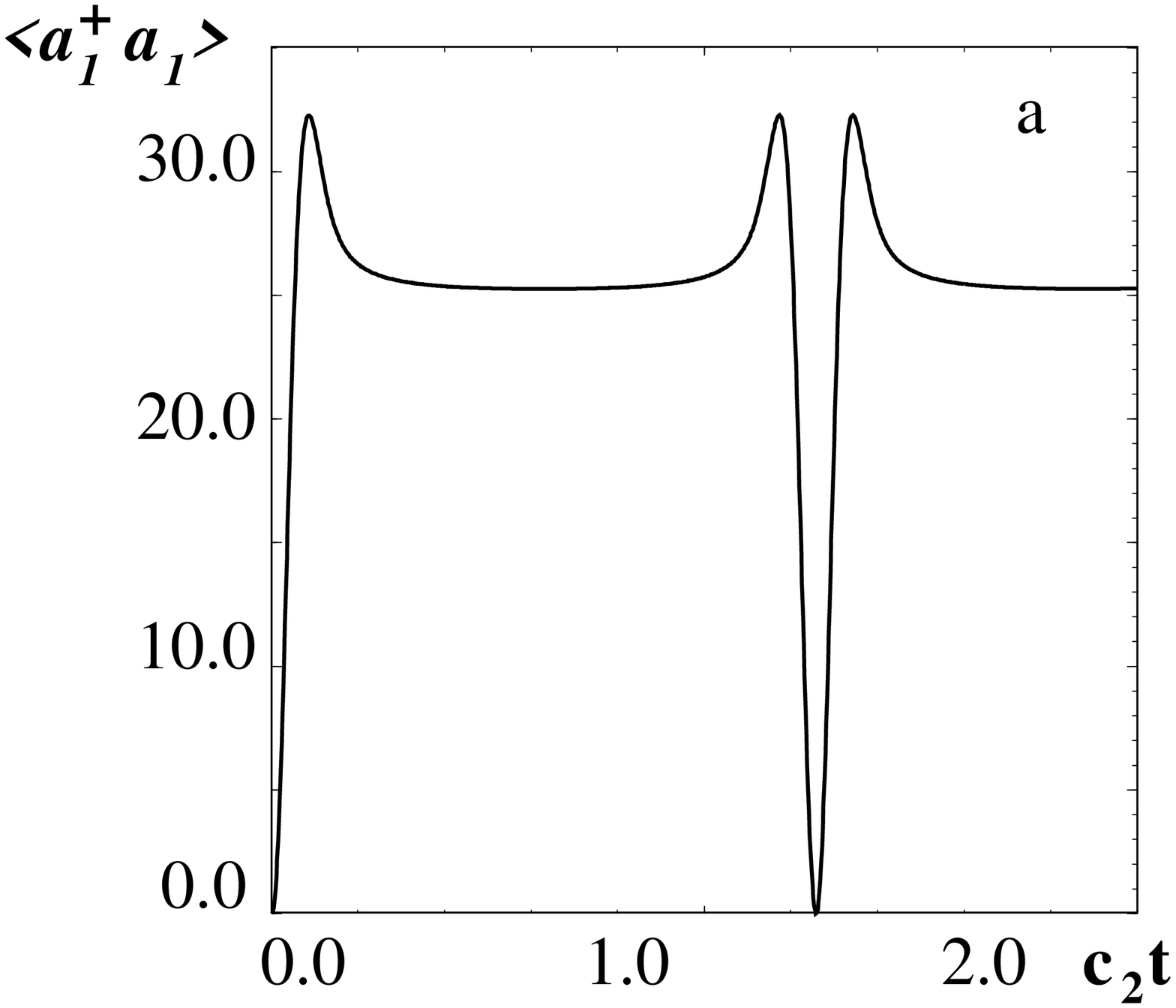,width=3.in}
\end{figure}
\begin{figure}
\epsfig{figure=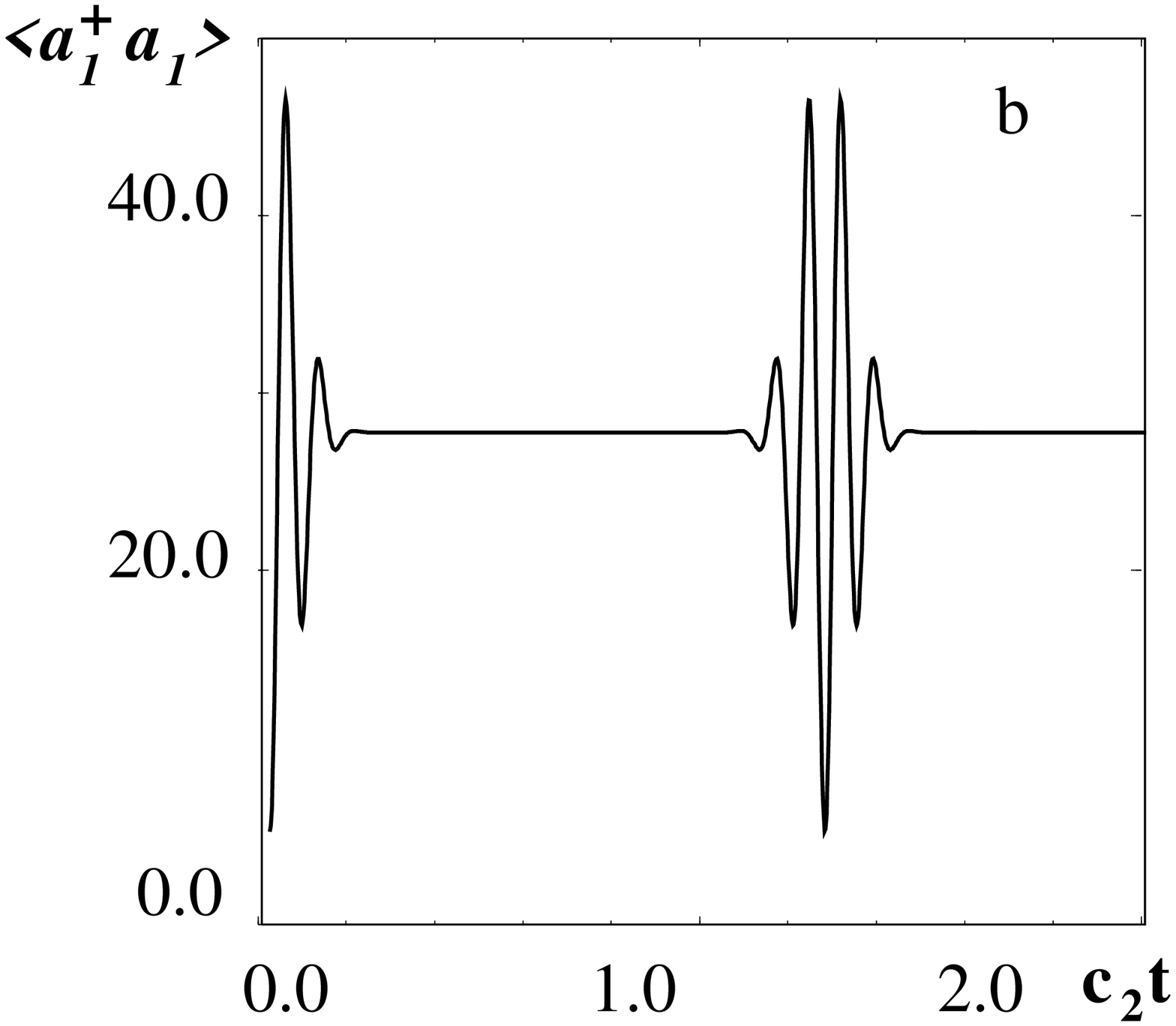,width=3.in}
\caption{
Time evolution of the side-mode population $\langle a_1^\dagger a_1\rangle$
for N=100 and (a)m=0, (b) m=5}
\end{figure}

\newpage
\begin{figure}
\epsfig{figure=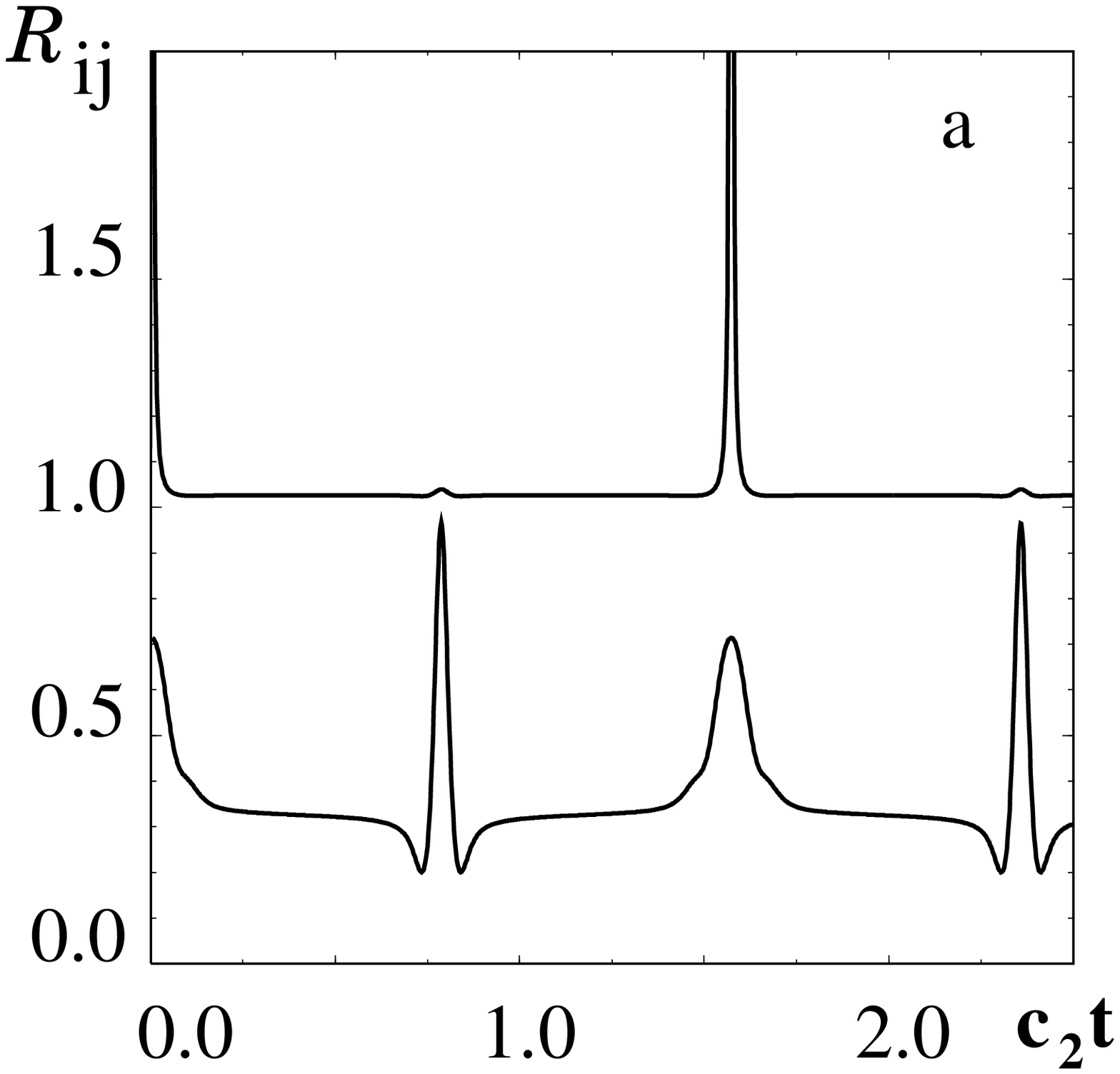,width=3.in}
\end{figure}

\begin{figure}
\epsfig{figure=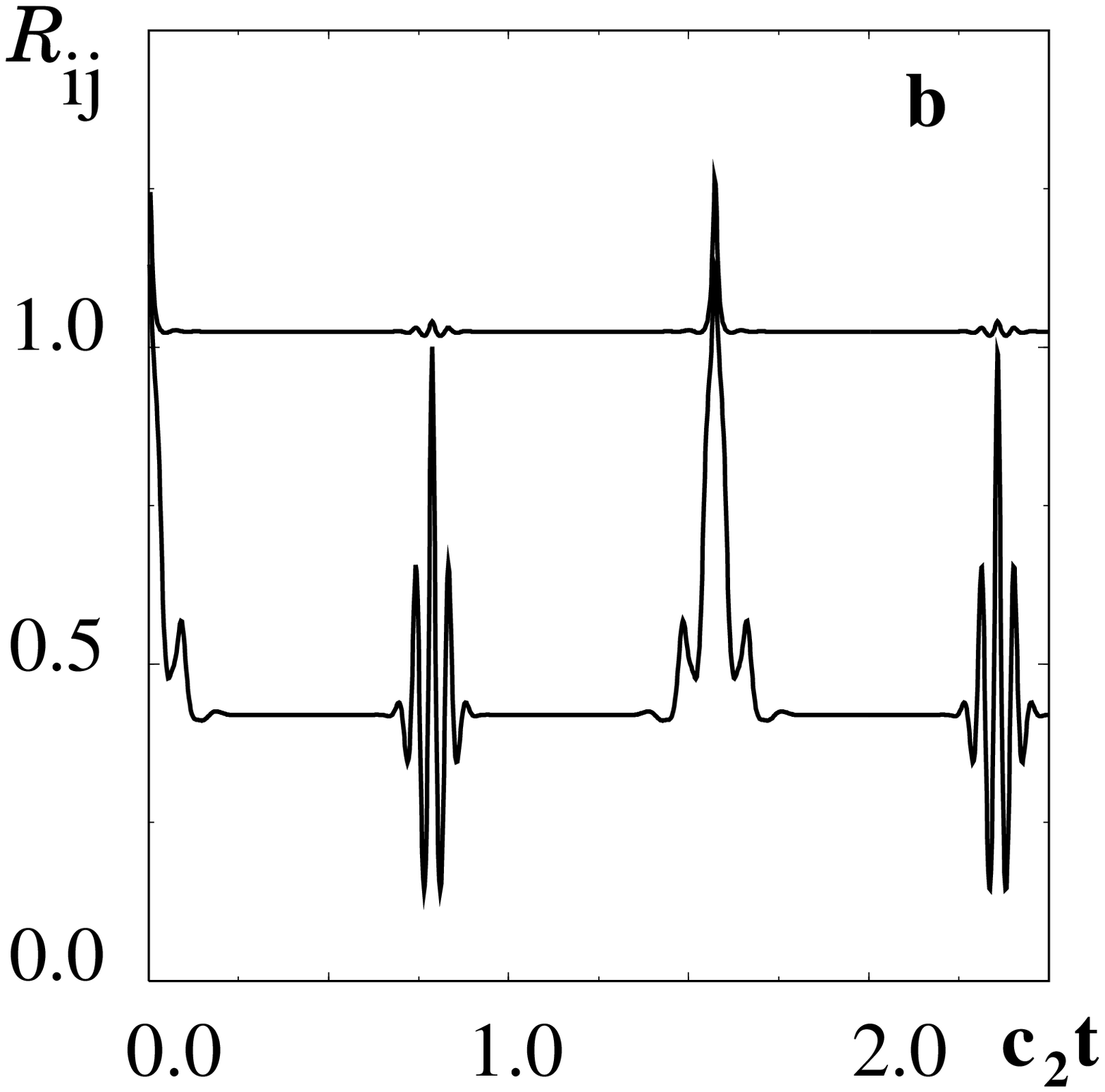,width=3.in}
\caption{Time evolution of the one-time normalized central-mode---side-mode
correlation function (lower curve) ${\cal R}^{(2)}_{01,1}(t)$
and the  one-time normalized side-mode---side-mode correlation function 
(upper curve) ${\cal R}^{(2)}_{-1,1}(t)$ for N=100 and (a)m=0, (b) m=5}
\end{figure}

\end{document}